\documentstyle[pra,aps,amsfonts,amssymb,graphics,epsfig,color]{revtex}

\begin{document}

\title{\LARGE\bf Stochastic collective dynamics of charged--particle
beams\\
in the stability regime\\ \vspace{15pt} \small(revision day :
September 27, 2000)}

\author{{\bf Nicola Cufaro Petroni$^*$,
Salvatore De Martino$^\S$, Silvio De Siena$^\S$} and {\bf Fabrizio
Illuminati$^\S$}}

\address{\vspace{8pt}
$^*$Dipartimento Interateneo di Fisica dell'Universit\`a e del
Politecnico di Bari; \\ INFN Sezione di Bari and INFM Unit\`a di
Bari;\\ Via G. Amendola 173, 70126 Bari, Italy;\\ Email:
cufaro@ba.infn.it; Web:
http://www.ba.infn.it/$\sim$cufaro/homepage.html\\
\vspace{5pt}$^\S$Dipartimento di Fisica dell'Universit\`a di
Salerno;\\ INFM Unit\`a di Salerno and INFN Sezione di Napoli,
Gruppo collegato di Salerno;
\\ Via S. Allende, I--84081 Baronissi (SA), Italy;\\ Email:
demartino@sa.infn.it, desiena@sa.infn.it, illuminati@sa.infn.it}

\maketitle \vspace{1cm}

\begin{abstract}
We introduce a description of the collective transverse dynamics
of charged (proton) beams in the stability regime by suitable
classical stochastic fluctuations. In this scheme, the collective
beam dynamics is described by time--reversal invariant diffusion
processes deduced by stochastic variational principles (Nelson
processes). By general arguments, we show that the diffusion
coefficient, expressed in units of length, is given by
$\lambda_c\sqrt{N}$, where $N$ is the number of particles in the
beam and $\lambda_c$ the Compton wavelength of a single
constituent. This diffusion coefficient represents an effective
unit of beam emittance. The hydrodynamic equations of the
stochastic dynamics can be easily recast in the form of a
Schr\"odinger equation, with the unit of emittance replacing the
Planck action constant. This fact provides a natural connection to
the so--called ``quantum--like approaches'' to beam dynamics. The
transition probabilities associated to Nelson processes can be
exploited to model evolutions suitable to control the transverse
beam dynamics. In particular we show how to control, in the
quadrupole approximation to the beam--field interaction, both the
focusing and the transverse oscillations of the beam, either
together or independently.
\end{abstract}

\vspace{1cm}\noindent PACS numbers: 02.50.Ey, 29.27.Bd, 41.75.Lx.

\section{Introduction}\label{introduction}

Most of the studies on the dynamics of charged beams in particle
accelerators are concerned with classical phenomena of nonlinear
resonances as isolated sources of unstable behaviors. Following
this line of thought, a general understanding of classical
dynamical processes in particle accelerators has been reached in
recent years \cite{conte}. However, the transverse coherent
oscillations of the beam density and profile require, to be
explained, some mechanism of local correlation and loss of
statistical independence. This implies the need to treat all the
interactions as a whole, and to introduce an effective collective
dynamics. Moreover, the overall interactions between charged
particles and machine elements are really nonclassical in the
sense that out of the many sources of noise that are present,
almost all are mediated by fundamental quantum processes of
emission and absorption of photons. Therefore the collective
effective descriptions of these processes could contain, in
principle, some quantum signature \cite{qabp1}.

Starting from the above considerations, different approaches to
the collective dynamics of charged beams have been developed. Some
of them, relying on the Fokker--Planck equation and the
statistical effects on the dynamics of colliding beams, have
become an established reference in treating the sources of noise
and dissipation in particle accelerators by standard classical
probabilistic techniques~\cite{ruggiero}. Other, more recent,
approaches are instead based on suitable coarse grainings of the
constitutive kinetic equations, and yield Schr\"odinger--like
equations, with a thermal unit of emittance replacing the Planck
action constant (``quantum--like approaches'' to beam
dynamics)~\cite{fedele}.

In this paper we first of all exploit (classical) mechanical
criteria of stability in order to establish a connection with the
statistical fluctuations affecting the beam dynamics. In
particular we deduce a phenomenological expression of the
characteristic unit of action (emittance) which quantifies the
amount of these fluctuations and which is ultimately related to
the microscopic scales and to the number of particles. Explicitly,
the resulting expression for the diffusion coefficient ${\cal{E}}$
in units of length (i.e. the transverse unit of emittance) turns
out to depend both on the fundamental Compton wavelength
$\lambda_{c}$ and on the number $N$ of particles constituting the
beam through the simple, but nontrivial, formula ${\cal{E}} =
\lambda_{c}\sqrt{N}$. Thus motivated, we model the collective beam
dynamics by introducing suitable stochastic processes with
long--range coherent correlations.

This kind of analysis considers the regime of stability of our
dynamical systems. In this framework we study the intermediate,
but physically relevant, regime of beam dynamics in which a
balance is realized, on the average, between the energy
dissipation and the external RF energy pumping. Therefore our
approach differs crucially from the previous ones, based only on
the Fokker--Planck equation, since in this regime the overall
classical beam dynamics can be considered at the same time
stochastic and time--reversal invariant.

This scenario hints to very interesting perspectives for the
following reasons. First of all, {\it classical} stochastic
dynamical systems with time--reversal invariance have been
introduced and extensively studied in the context of Nelson
stochastic mechanics~\cite{nelson}. They are by now a well
understood subject, both from the physical and mathematical point
of view. For instance, among the possible physical applications of
this modelling it is worth noticing the recovering of the
Titius--Bode law for planet orbits in the solar system obtained in
reference~\cite{albeverio}. The study of these dynamical systems
is based on an extension of the variational principles of
classical mechanics to include the case of a diffusive kinematics
replacing the deterministic one~\cite{guerravariaz}. This is
remarkable since variational principles are a very powerful tool
in the description of physical systems. In the present case the
stochastic variational principle yields two coupled hydrodynamic
equations, respectively for the density and for the forward
velocity field, which provide an effective description of the
transverse oscillations of the beam density in the regime of
stability.

On the other hand it is also interesting to remark that the two
nonlinear coupled hydrodynamic equations of the stochastic
mechanics are equivalent to one linear equation of the form of a
Schr\"odinger equation, with the Planck action constant replaced
by the diffusion coefficient of the random kinematics. This fact
connects our approach to the quantum--like approaches to beam
dynamics. Moreover, since this description involves not only a
Fokker--Planck equation but also a dynamical prescription
connected with an external potential, it allows to implement the
powerful techniques of active control for the dynamics of the
beam. This is at variance with the case of a purely dissipative
Fokker--Planck dynamics which only describes a passive
(irreversible) evolution of the state, and where you have no
control whatsoever on the velocity field.

In fact, on the basis of the description of the beam collective
dynamics in terms of the hydrodynamic equations of Nelson
stochastic mechanics with the proper diffusion coefficient, we
will show how we can implement techniques of control already
developed in the general context of stochastic dynamical
systems~\cite{cufarojpa}. These techniques exploit the transition
probabilities, a fundamental object in the theory of diffusion
processes, in order to drive the beam toward a specified and
controlled evolution. In particular in our quantum--like approach
we construct time--dependent potentials which drive the system
toward final states characterized by an improved collimation. At
the same time, and independently, also the transverse betatron
oscillations can be controlled and varied.

The paper is organized as follows: In Section~\ref{calogerism} we
exploit some basic criteria of mechanical stability in order to
supply a phenomenological support for our fluctuative approach to
collective dynamics of beams in particle accelerators. In
particular these criteria will allow to connect the (transverse)
emittance to the characteristic microscopic scale and to the total
number of the particles in a bunch. In Section~\ref{nelsonism} we
introduce a time--reversal invariant, stochastic description of
the collective dynamics of the beam in the stability regime. The
(hydrodynamic) equations of motion for the density and the profile
of a bunch are here derived from a stochastic variational
principle. In Section~\ref{control} we sketch the general
structure of controlled dynamics for quantum and quantum--like
systems. In Section~\ref{squeezing} we explicitly construct, as
mentioned above, some examples of controlled beam evolution in the
quadrupole approximation to the beam--field interaction. Finally
in Section~\ref{conclusions} conclusions follow.

\section{Collective behavior of dynamical systems
in the stability regime}\label{calogerism}

Effective wave equations of the Schr\"odinger form, but associated
to a non fundamental unit of action (that is, different from the
Planck constant), have been introduced to describe the collective
dynamics for classical physical systems with many degrees of
freedom, including optical fibers~\cite{ofib} and charged particle
beams in accelerators~\cite{fedele}. The basic feature common to
all these systems is their high degree of coherence, which allows
to introduce an effective description in terms of collective
degrees of freedom. These represent the cooperative dynamical
behavior of the many constituents of the system. This collective
motion is summarized by the effective equations for the
configurational (or phase--space) density and for the velocity
fields.

Here we specialize to the dynamics of charged beams a general
scheme previously introduced for the study of the dimensions of
stability for macroscopic and mesoscopic systems~\cite{demartino}.
The ultimate goal of the analysis of stability is to single out a
(minimal) unit of action in terms of the characteristic constants,
of the form of the interaction, and of the linear dimensions of
the considered system. This analysis will allow us to relate the
parameters associated to the collective degrees of freedom with
the characteristics of the microscopic constituents. In
Appendix~\ref{minimalaction} we summarize the procedure, while
here we will just quickly quote the results.

We introduce a unit of action $\alpha$ (which will turn out to be
minimal) defined by:
\begin{equation}\label{action}
\alpha=m {\tilde v}^2\tau,
\end{equation}
where ${\tilde v}$ denotes the characteristic mean velocity per
particle in the system, while $\tau$ is a characteristic
microscopic time whose size must be determined self--consistently.
Imposing suitable criteria of stability, we obtain as first
result~\cite{demartino} that the order of magnitude of this small
time $\tau$ in~(\ref{action}) must be given by
\begin{equation}\label{timescale}
\tau \cong \frac{{\cal T}}{\sqrt{N}},
\end{equation}
where  ${\cal{T}}$ is the macroscopic time scale associated to the
entire system, and defined through the relation ${\tilde
v}=R/{\cal{T}}$, where $R$ denotes the global length scale of the
system. Therefore ${\cal{T}}$ has the meaning of a characteristic
traveling time for a particle inside the system. Moreover, we
obtain for the (minimal) unit of action $\alpha$ the general
expression:
\begin{equation}\label{force}
\alpha\cong m^{1/2}R^{3/2} \sqrt{F(R)},
\end{equation}
where $F(R)$ denotes the value of the force ruling the system,
computed on a distance scale of the order of magnitude of the
global scale of the system~\cite{demartino}.

It must be noted that the relation~(\ref{timescale}) has been
originally conjectured by F. Calogero~\cite{calogero}  in a
different context. It is remarkable that, when tested in the range
of all known stable macroscopic and mesoscopic aggregates of
particles, the relation~(\ref{force}) always yields the order of
magnitude of the Planck action constant~\cite{demartino}. This is
the reason why we identify $\alpha$ as a minimal unit of action.
It was not obviously trivial that from a purely mechanical
criteria of stability the fundamental microscopic scale of action
could emerge. Moreover this fact, together with the observation
that we have self--consistently obtained the
expression~(\ref{timescale}) for the microscopic time $\tau$,
strongly hints to two further conclusions.

First, the factor $1/\sqrt{N}$ in the scaling
relation~(\ref{timescale}) typically hints to the presence of
collective fluctuations, whose characteristic scale of time is
given by $\tau$. This is not a surprising fact due to the large
number of constituents in the system. Second, the universal
coincidence of the minimal unit of action with the Planck action
constant strongly points to the fact that these collective
fluctuations are ultimately connected with the fundamental
microscopic scales. It is worth noting that, as we will show
later, this fact is not connected to complicated or mysterious
effects of direct quantum origin, but it simply takes into account
the constraint given by the characteristic spatial extension of
the microscopic constituents.

In the specific instance of charged beams, we first verify the
numerical coincidence of $\alpha$ with the Planck action constant.
We then single out the expression of the (transverse) emittance in
terms of the microscopic minimal length scale, and of the number
of elementary constituents. This second step of our analysis is
performed in the particular instance of proton beams since it
would be impossible to find a classical characteristic length
extension for the electron.

In the first step we consider a representative proton (electron),
in the reference frame comoving with the bunch. Confinement and
stability for the transverse motion of the bunch arise from the
many interactions both among its constituents and between the same
constituents and the external focusing electromagnetic fields. It
is well known that the net effect can be, in the first quadrupole
approximation, summarized by a harmonic force of modulus $F(r)
\cong Kr$, where $K$ is the effective phenomenological elastic
constant associated to the transverse dynamics. Then
equation~(\ref{force}) yields
\begin{equation}\label{beamaction}
\alpha\cong m^{1/2} R^{2} K^{1/2}.
\end{equation}
We can now estimate $\alpha$ by introducing, besides the proton or
the electron mass, the experimental values for the transverse
linear dimension $R$ and for the effective elastic constant $K$.
We have~\cite{conte} $K  \cong 10^{-12}Nm^{-1}$ (transverse
oscillations of protons at Hera) and $K \cong 10^{-11}Nm^{-1}$
(transverse oscillations of electrons in linear colliders), while
$R \cong 10^{-7}m$ in both experimental situations. As a
consequence, in both cases equation~(\ref{beamaction}) yields
$\alpha \cong h$. We have therefore reached our first goal.

We now move in the second step to single out the parameter
associated to the stability of the system at the mesoscopic scale
in the case of charged beams. This parameter is given in terms of
a characteristic unit of (transverse) emittance. The emittance is
a scale of action that measures the spread of the bunch in phase
space. It can also be defined as a unit of equivalent temperature
or, in configuration space, as a unit of length. It is clear that
this quantity must depend on the characteristic scales and on the
total number of the elementary components in the system. In the
framework of our scheme we are able to provide, at least in order
of magnitude, a quantitative estimate of this dependence.

We proceed as follows: in the regime of stability and of thermal
equilibrium, that we explicitly consider, the emittance can be
expressed as a unit of equivalent thermal action. We denote by $T$
the equivalent unit of equivalent temperature of the system
(namely the unit of energy divided by the Boltzmann constant
$k_B$), and we define the characteristic thermal unit of action
associated to the system as $k_B T{\cal{T}}$, the product of the
unit of thermal energy and of the characteristic global time. In
our scheme, when $\alpha\cong \hbar$, the time $\tau$ connected to
the microscopic scales can also be identified with the usual scale
of time associated to a microscopic system at the equilibrium
temperature $T$; hence
\begin{equation}\label{temperature}
\tau \cong \frac{h}{k_{B}T}\,.
\end{equation}
Using relation~(\ref{timescale}) we finally obtain the equivalent
thermal unit of action, the transverse emittance $\epsilon$, in
terms of the minimal action $\hbar$ and of the total number of
particle $N$:
\begin{equation}\label{transvemittance}
\epsilon \equiv \frac{k_{B}T{\cal T}}{2 \pi} \cong \hbar \sqrt{N}
\, .
\end{equation}

Up to now our results hold both for protons and electrons. However
the previous relation allows a more direct check if written in
terms of characteristic units of length. We know however that
electrons do not possess a finite characteristic length extension,
while for protons we know that such a linear extension coincides,
in order of magnitude, with the Compton wave length. Thus,
specializing to protons, we can show that, at least in order of
magnitude, the numerical value of the transverse dimension of the
bunch is:
\begin{equation}\label{transvdimension}
\frac{\epsilon}{mc} \cong \lambda_c \sqrt{N},
\end{equation}
where the Compton wavelength $\lambda_{c} = \hbar /mc$ ($m$ is the
proton mass, and $c$ is the velocity of light).

We can now interpret equation~(\ref{transvdimension}) in the
following, simple way: the (transverse) mean dimension of the beam
at equilibrium is connected to the characteristic length scale
$\lambda_{c}$ of its microscopic constituents through the scaling
factor $\sqrt{N}$. This last peculiar form, in turn, suggests a
fluctuation mechanism which stabilizes the system. As previously
anticipated, the microscopic scales influence the system only
through the minimal length scale, i.e. the length extension of the
elementary constituents, without direct connections to more
involved quantum effects.

Inserting in equation~(\ref{transvdimension}) the numerical data
of the proton wave length and of the number of proton in typical
accelerators~\cite{conte}, we obtain the experimental order of
magnitude of the transverse dimension of the bunch
$10^{-7}\div10^{-8}\, m$~\cite{demartino}. On the basis of our
phenomenological scheme, we introduce in the next section a
quantitative stochastic description of beam dynamics in the
stability regime. In the following, consistently with the analysis
carried out so far, we take as diffusion coefficient, expressed in
unit of length, for the stochastic kinematics the quantity
(transverse emittance)
\begin{equation}\label{emittance}
{\cal E} \equiv \frac{\epsilon}{2mc} \cong \frac{\lambda_c}{2}
\sqrt{N} \, ,
\end{equation}
where the factor $2$, which does not affect the order of
magnitude, is introduced for later computational convenience.

\section{Stochastic collective dynamics in the stability
regime}\label{nelsonism}

In this section we model the spatial fluctuations (associated to
the diffusion coefficient~(\ref{emittance})) via the random
kinematics performed by a representative particle that oscillates,
in a reference frame comoving with the bunch, around the closed
ideal orbit. This representative particle is identified with the
collective degree of freedom by letting the associated probability
density coincide with the real density of particles in the bunch.
This last step is achieved by suitably rescaling the normalization
of the total number of particles. Before proceeding, we establish
the notations according to the standard conventions.

We denote ${\bf r} \equiv (x, y)$ a point in the transverse
section orthogonal to the beam direction. We then measure the time
in unit of length through the arc length $s$ along the design
orbit (curvilinear coordinate). We now consider the
(two--dimensional) diffusion process ${\bf q} (s)$ which describes
the motion of the representative particle and whose probability
density coincides with the particle density of the bunch in the
transverse direction. The evolution in the ``time'' $s$ of the
process ${\bf q}$ is described by the It\^o stochastic
differential equation
\begin{equation}\label{ito}
 d{\bf q}(s) = {\bf v}_{(+)}({\bf q}(s), s)ds +
\sqrt{\cal{E}}d{\bf w} (s) \, ,
\end{equation}
where ${\bf v}_{(+)}$ is the (forward) drift, $d{\bf w} (s) \equiv
{\bf w} (s + ds) - {\bf w} (s)$ is the $\delta$--correlated time
increment of the standard Wiener noise, and, as already
anticipated, the diffusion coefficient is the characteristic
transverse emittance. Equation~(\ref{ito}) defines the random
kinematics performed by the collective degree of freedom.

In the stability regime the energy lost by photonic emissions is
regained in the RF cavities, and on average the dynamics is
time--reversal invariant. We are thus in a situation in which
there are both a random kinematics and time reversal invariance.
Therefore the dynamics must be independently added to the
kinematics (at variance with the purely dissipative Fokker--Planck
case) by introducing a suitable stochastic least action
principle~\cite{guerravariaz}. The latter is obtained as a
generalization of the variational principle of classical
mechanics, by replacing the classical deterministic kinematics,
$d{\bf q}_{c}(s) = {\bf v}_{c}(s) ds$, with the random diffusive
kinematics of equation~(\ref{ito}). The equations of motion thus
obtained take the form of two coupled hydrodynamic equations
describing the evolution in time of the beam density and of the
velocity field of the beam profile. In the following we give a
brief sketch of the stochastic variational method, and we
introduce the coupled hydrodynamic equations, referring for
details to reference~\cite{guerravariaz}.

Given the stochastic differential equation~(\ref{ito}), one can
associate to the diffusion process ${\bf q}(s)$ a probability
density $\rho({\bf r}, s)$, where ${\bf r} \equiv (x, y)$ denotes
the transverse coordinates (the radial coordinate and the vertical
coordinate). Besides the forward drift ${\bf v}_{(+)}({\bf r},
s)$, we can define a backward drift ${\bf v}_{(-)}({\bf r}, s)  =
{\bf v}_{(+)}({\bf r}, s) - 2 {\cal E} (\nabla \rho) ({\bf r}, s)
/ \rho ({\bf r}, s)$, with $\nabla \equiv (\partial_{x},
\partial_{y})$. It is useful to introduce two new
variables, ${\bf v}({\bf r}, s)$ and ${\bf u}({\bf r}, s)$,
respectively the current and the osmotic velocity fields, defined
as:
\begin{equation}\label{osmotic}
{\bf v}=\frac{{\bf v}_{(+)} + {\bf v}_{(-)}}{2} \; ;\qquad\quad
{\bf u}=\frac{{\bf v}_{(+)}-{\bf v}_{(-)}}{2}\, =\,
{\cal{E}}\frac{\nabla \rho}{\rho} \, .
\end{equation}
The velocities in equation~(\ref{osmotic}) have a transparent
physical meaning: the current velocity ${\bf v}$ represents the
global velocity of the density profile, being the stochastic
generalization of the velocity field of a classical perfect fluid.
On the other hand the osmotic velocity $\bf u$ is clearly of
intrinsic stochastic nature, for it is a measure of the non
differentiability of the stochastic trajectories, and it is
related to the spatial variations of the density.

In order to establish the stochastic generalization of the least
action principle, one introduces a mean classical action in strict
analogy to the classical deterministic action. The main difficulty
in the stochastic case is due to the non differentiable character
of the sample paths of a diffusion process which does not allow to
define the time derivative ${\dot {\bf q}}$ of the process. Such a
definition is possible only in an average sense trough a suitable
limit on expectations. The stochastic action is then defined
as~\cite{guerravariaz}
\begin{equation}\label{lagrangianaction}
A(s_{0}, s_{1}; {\bf q(\cdot)}) = \int_{s_{0}}^{s_{1}}
\lim_{\Delta s \rightarrow 0^{+}} {\bf E} \left[ \frac{m}{2}
\left( \frac{\Delta {\bf q}}{\Delta s} \right)^{2} - V({\bf q})
\right] ds \, ,
\end{equation}
where ${\bf E}(\,\cdot\,) = \int dr (\,\cdot\,) \rho ({\bf r}, s)$
denotes the expectation of functions of the process with respect
to the probability density, $V$ denotes an external potential, and
$\Delta {\bf q}(s) = {\bf q}(s + \Delta s) - {\bf q}(s)$. It can
be shown that the mean action~(\ref{lagrangianaction}) associated
to the diffusive kinematics~(\ref{ito}) can be recast in the
following particularly appealing Eulerian hydrodynamic form
\cite{nelson}:
\begin{equation}\label{eulerianaction}
A(s_{0}, s_{1}; {\bf v},\rho) = \int_{s_{0}}^{s_{1}}ds\int d{\bf
r} \left[ \frac{m}{2} \left( {\bf v}^{2} - {\bf u}^{2} \right) -
V({\bf r}) \right] \rho({\bf r}, s) \, ,
\end{equation}
where ${\bf v}$ and ${\bf  u}$ are defined in
equation~(\ref{osmotic}). The stochastic variational principle now
follows by imposing the stationarity of the stochastic action
($\delta A = 0$) under smooth and independent variations $\delta
\rho$ of the density, and $\delta {\bf v}$ of the current
velocity, with vanishing boundary conditions at the initial and
final times.

As a first consequence we get that the current velocity has a
gradient form:
\begin{equation}\label{gradient}
m{\bf v} ({\bf r}, s) = \nabla S ({\bf r}, s) \, ,
\end{equation}
while the non linearly coupled Lagrange equations of motion for
the density $\rho$, and for a current velocity ${\bf v}$ of the
form~(\ref{gradient}) are: the continuity equation typically
associated to every diffusion process
\begin{equation}\label{continuity}
\partial_{s} \rho = -\nabla \cdot (\rho {\bf v})\,,
\end{equation}
and a dynamical equation
\begin{equation}\label{hjm}
\partial_{s} S + \frac{m}{2} {\bf v}^{2} - 2m {\cal{E}}^2
\frac{\nabla^{2} \sqrt{\rho}}{\sqrt{\rho}} + V({\bf r}, s) = 0 \,,
\end{equation}
which characterizes the particular class of time--reversal
invariant diffusion processes (Nelson processes). Last equation
has the same form of the Hamilton--Jacobi--Madelung (HJM)
equation, originally introduced in the hydrodynamic description of
quantum mechanics by Madelung~\cite{madelung}. It can also be
shown that the continuity equation~(\ref{continuity}) is
equivalent to the standard Fokker--Planck equation
\begin{equation}\label{fp}
\partial_{s} \rho = -\nabla \cdot [{\bf v}_{(+)} \rho]
+  {\cal E} \, \nabla^2  \rho  \, ,
\end{equation}
by simple substitution from~(\ref{osmotic}). The time--reversal
invariance is assured by the fact that the forward drift velocity
${\bf v}_{(+)} ({\bf r}, s)$ is not a field given {\it a priori},
as usual for diffusion processes of the Langevin type; instead it
is dynamically determined at any instant of time, starting by an
initial conditions, through the HJM evolution
equation~(\ref{hjm}).

The equations~(\ref{continuity}) and~(\ref{hjm}) describe the
collective behaviour of the bunch at each instant of time through
the evolution of both the particle density and the velocity field
of the bunch. In particular we can calculate the expectations
${\mathbf E} [{\bf q} (s)]$ and ${\mathbf E} [{\bf v} ({\bf q}
(s), s)]$, which supply the coordinates and the velocity
components of the center of the bunch profile at time $s$, while
the variances ${\mathbf V} [q_i (s)] \equiv \sqrt{{\mathbf E}
[q_i^2 (s)] - {\mathbf E}^2 [q_i (s)]}$ represent the spreading of
the bunch density along each space direction.

It is finally worth noticing that, introducing the trivial
representation~\cite{madelung}
\begin{equation}\label{ansatz}
\psi ({\bf r}, s) = \sqrt{\rho({\bf r}, s)}\, {\mathrm e}^{i
S({\bf r}, s)/2 m {\mathcal E}} \, ,
\end{equation}
the coupled equations~(\ref{continuity}) and~(\ref{hjm}) are
equivalent to a single linear equation of the form of the
Schr\"odinger equation in the function $\psi$, with the Planck
action constant replaced by the emittance ${\cal E}$:
\begin{equation}\label{schroed}
  i2m{\mathcal E}\partial_s\psi=-2m{\mathcal
  E}^2\nabla^2\psi+V\psi\,.
\end{equation}
In this formulation the ``wave function'' $\psi$ carries the
information on both the dynamics of the bunch density $\rho$, and
of the velocity field of the bunch, where the velocity field is
determined through equation~(\ref{gradient}) by the phase function
$S({\bf r}, s)$. This shows, as previously claimed, that our
procedure, starting from a different point of view, leads to a
description formally analogous to that of the quantum--like
approaches to beam dynamics~\cite{fedele}.

\section{Construction of controlled states for quantum and quantum--like
systems}\label{control}

In the previous section we have introduced two coupled equations
that describe the dynamical behaviour of the beam: the first is
the It\^o equation~(\ref{ito}), or equivalently the Fokker-Planck
equation~(\ref{fp}); the second is the HJM equation~(\ref{hjm}).
Here, we briefly sum up (with the present notations) a general
procedure exploited in reference~\cite{cufarojpa} to control the
dynamics of quantum and quantum-like systems, while in the next
section we will give an explicit application of the method to the
transverse beam dynamics. From now on we will consider
one--dimensional processes denoting by $\xi$ a one--dimensional
space variable, in suitable units. In the next section the
variable $\xi$ will be one of the transverse space coordinates. In
reference~\cite{cufarojpa}, it has been shown that given a pair of
functions $\rho(\xi, s)$ and $v_{(+)}(\xi, s)$ (density and
forward velocity) which satisfy~(\ref{fp}) or
equivalently~(\ref{continuity}), the equation~(\ref{hjm}) with the
given functions allows one to compute a control potential $V_c$.
Remark that $\rho(\xi, s)$ and $v_{(+)}(\xi, s)$ can also be an
entire class of functions of a given form.

Let us take for instance the solution $\rho (\xi, s)$ of a
Fokker--Planck equation~(\ref{fp}) with a given $v_{(+)}(\xi, s)$
and a constant diffusion coefficient ${\cal E}$, define the
function $W(\xi, s)$ from
\begin{equation}\label{RW}
m v_{(+)}(\xi, s) = \partial_{\xi}
W(\xi, s) \, ,
\end{equation}
and remind from~(\ref{osmotic}) and~(\ref{gradient}) that the
relation
\begin{equation}\label{RS}
m v_{(+)} = \partial_{\xi}
\left(S+{\cal E}\ln\tilde \rho \right)
\end{equation}
must hold, where $\tilde \rho$ is the adimensional function
(argument of a logarithm) obtained from the probability density
$\rho $ by means of a suitable and arbitrary multiplicative
constant with the dimensions of $\xi$. Hence from~(\ref{RW})
and~(\ref{RS}) we obtain for the phase function
\begin{equation}\label{phase}
S(\xi, s) = W(\xi, s) - m{\cal E} \ln \tilde \rho (\xi, s) -
\theta (s) \, ,
\end{equation}
which allows to determine $S$ from $\rho$ and $v_{(+)}$ up to an
additive arbitrary function of time $\theta (s)$. The functions
$\rho$ and $S$, satisfying our kinematical relations~(\ref{fp}),
are a solution of our dynamical problem if they also satisfy the
HJM equation~(\ref{hjm}). Since $S$ and $\rho$ are now fixed, this
equation must be considered as a (constraint) relation defining a
controlling potential $V_{c}$ which, after straightforward
calculations, turns out to be of the form:
\begin{equation}\label{potential}
V_{c}(\xi, s)=m {\cal E}^2\,
\partial_{\xi}^2\ln \tilde\rho +
m {\cal E} \left(\partial_s \ln \tilde \rho + v_{(+)}
\partial_{\xi} \ln \tilde \rho \right) - {mv_{(+)}^2\over2}
- \partial_s W + \dot\theta(s) \, .
\end{equation}
When the density $\rho$ interpolates between an initial and a
final distribution, then the controlling potential $V_c$
interpolates between the corresponding initial and final
potential. It is worth noticing that for a class of velocities
$v_{(+)}$ (i.e. the non singular, time--independent velocities,
but also particular instances of time--dependent velocities) the
Fokker--Planck equation~(\ref{fp}) alone would drive the density
towards an asymptotic solution which does not depend on the
initial condition (for details see~\cite{cufarojpa}
and~\cite{cufaropla}): this kind of evolution is not controlled by
an external potential. In the time--independent case the
asymptotic solution is also a stationary one. On the other hand,
when the Fokker--Planck equation~(\ref{fp}) is coupled with the
dynamical HJM equation~(\ref{hjm}) we have a way to control the
evolution and the right potential has the form~(\ref{potential})
which depends on the velocity $v_{(+)}$. This method can in
principle be applied to very complicated systems: for instance in
the beam dynamics we could keep the beam coherent even in the
presence of aberrations. However this problem is non explicitly
solvable in closed form and requires some approximate treatment.
At present we consider only the more simple, but still nontrivial,
case of the quadrupole approximation to the beam--field
interaction. In this case we can exactly compute controlling,
quadratic potentials which drive the bunch to a final state with
better focusing. Moreover we can avoid a technical difficulty
present in the more general situation. Actually the general
procedure often implies an initial singular behaviour in the phase
function. In fact, when we suddenly impose to the initial state
the forward drift associated to the final state, the new phase
turns out to be ``wrong'' with respect to the initial density.
Hence a ``kick'' in the potential is needed in order to produce
such a sudden change in the phase. This fact shows that to be
physically meaningful our procedure requires some smoothing.
In~\cite{cufarojpa}, however, it was noted that, at least for a
Gaussian choice of the initial and final densities, it is
particularly simple to implement transitions which do not need any
smoothing procedure. We can adopt this especially simple solution
exactly in the case of beam dynamics in the quadrupole (harmonic
potential) approximation.

\section{Controlled beam dynamics in the quadrupole
approximation}\label{squeezing}

We now move on to construct explicit examples of controlled beam
dynamics. In considering an accelerating machine we assume, as
usual, that the longitudinal and the transverse dynamics can be
deemed independent with a high degree of approximation. We will
work in the framework of the quadrupole approximation, with the
further simplification of considering decoupled evolutions along
the radial direction $x$ and the vertical direction $y$ in the
local reference frame.

Under these conditions, we can split the original,
two--dimensional diffusion process into two independent,
one--dimensional processes respectively along $x$ and $y$, each
ruled by a harmonic potential. The configurational variable $\xi$
of the previous section can here indifferently be either $x$ or
$y$ depending on the considered transverse direction. The
potential in each transverse direction will have the general form:
\begin{equation}\label{harmonic}
V(\xi, s) = \frac{1}{2} m \omega^2 (s) \xi^2 -m f(s)\xi+m U(s)\, .
\end{equation}
We have considered here a time--dependent frequency (parametric
oscillator) in order to describe also the effects due to strong
focusing~\cite{conte}. Note that here we have a potential measured
in units of mass, consistently with the choice of measure units
made in the Section~\ref{calogerism}. Our aim is now to exploit
the hydrodynamic equations~(\ref{continuity}) and~(\ref{hjm}) as
control equations for the beam dynamics. In particular, we will
show how to compute a controlling, time--dependent potential which
allows to drive a bunch prepared in a state with a certain degree
of collimation towards a final state with better focusing.

We consider a Gaussian shape for the initial density profile of a
bunch in each transverse direction, with constant dispersion, and
with the centre of the profile which performs a classical harmonic
motion with the same frequency associated to the initial
potential~(\ref{harmonic}). The motion of the centre models the
betatron oscillations of the bunch. In our quantum--like approach,
the state of the bunch is thus formally represented by a coherent
state. As anticipated at the end of the previous section, we will
now consider an instance of controlled evolution that does not
require an extra smoothing procedure for the driving velocity
field, i.e. the transition between pairs of Gaussian densities. In
particular we will describe transitions from a coherent
oscillating packet to another Gaussian state with a better
collimation (smaller dispersion). It is worth noticing that we can
also implement a procedure that allows to vary independently the
dispersion (collimation) of the bunch density and the motion of
the centre of the density profile (characteristics of the betatron
oscillations).

To this end we will recall~\cite{bvth} that if the velocity field
of a Fokker-Planck equation~(\ref{fp}) with constant diffusion
coefficient ${\cal E}$ (the transverse emittance) has the linear
form $v_{(+)}(\xi, s) = A(s) + B(s)\xi$, with $A(s)$ and $B(s)$
continuous functions of $s$, then there are always Gaussian
solutions ${\cal N}\bigl(\mu(s), \nu (s) \bigr)$, where $\mu(s)$
(the displacement of the centre of the Gaussian distribution) and
$\nu (s)$ (the variance of the Gaussian distribution) are
solutions of the differential equations
\begin{equation}\label{parameters}
  \mu'(s) - B(s)\mu(s) = A(s)\,;\qquad
     \nu'(s) - 2 B(s) \nu(s) = 2 {\cal E} \, ,
\end{equation}
with suitable initial conditions, and where the prime denotes the
derivative with respect to $s$. As previously stated, all along
the time evolution our states keep a Gaussian shape for the
density, and the centre of the density profile performs an
arbitrarily assigned motion. Then, if we adopt the concise
quantum--like representation of the bunch state~(\ref{ansatz}) it
is straightforward to show that the general form for the wave
packet will be:
\begin{equation}\label{oscillating}
\psi(\xi, s)= \left(\frac{1}{2 \pi\nu}\right)^{1/4}
\exp\left[-{(\xi - \mu)^2 \over4\nu} +\frac{i}{2m{\cal E}}
\left(m\mu'\xi + m \frac{\nu'}{4\nu} (\xi - \mu)^2 + \theta
\right)\right] \, ,
\end{equation}
while the forward velocity field reads
\begin{equation}\label{oscillvelocity}
v_{(+)}(\xi, s) = \mu'+\frac{\nu'-2{\cal E}}{2\nu} (\xi-\mu)\,.
\end{equation}
Here the $s$-dependent functions $\mu (s)$ and ${\nu} (s)$
describe respectively the motion of the centre of the density
profile and the spreading of the bunch density in the chosen
transverse direction; on the other hand $\theta (s)$ plays the
role of an arbitrary integration constant as can be seen
from~(\ref{phase}). Of course a suitable potential must also be
tailored from the equation~(\ref{potential}) in order to keep the
evolution of the wave function~(\ref{oscillating}) on the right
track: we will show that in fact this control potential has the
form suggested in~(\ref{harmonic}).

The equation~(\ref{oscillating}) represents the most general
Gaussian packet, with a given generic motion $\mu (s)$ of its
centre and with a given dispersion $\nu (s)$, associated to a
linear form of the forward velocity in the Fokker--Planck
equation~(\ref{fp}). This also allows us to keep independent the
initial and the final motion of the centre of the packet from the
dispersion. As a first example let us now consider the transitions
between two states of the form~(\ref{oscillating}) with constant
dispersion and with a harmonic motion of the centre of the
profile. If initially (namely for $s\ll\tau$, where from now on
$\tau$ is the transition instant) we start with $\nu(s) = \nu_1$
and $\mu (s) = a_1\cos (\omega_{1} s)$, we will have an initial
Gaussian density profile with spreading $\nu_1$ and with harmonic
betatron oscillation of frequency $\omega_1 = {\cal E}/\nu_1$. We
now want to drive the system towards a final (for $s\gg\tau$)
state of the form~(\ref{oscillating}), but with a spreading $\nu_2
< \nu_1$ (better collimation) and a new betatron oscillation
$\mu_2 (s)$. To this end we only need to put in the solution
${\cal N} (\mu (s), \nu (s))$ two functions $\mu (s), \nu (s)$
which interpolate between the corresponding initial and final
functions of the motion of the centre, and of the spreading
respectively. Moreover, with a suitable choice of the
$\xi$--independent part of the phase function
in~(\ref{oscillating}), the forward velocity field will also
smoothly interpolate between the initial and the final velocity
fields~\cite{cufarojpa}. The control potential which drives the
solution toward the required end is finally obtained by the
equation~(\ref{potential}) with ${\tilde \rho}$ given by the
interpolating solution ${\cal N} (\mu (s), \nu (s))$, and with
$v_{(+)}$ given by the associated forward velocity. Of course
there is a large number of possible choices for the interpolating
functions $\mu (s), \nu (s)$: this will allow us to single out the
forms that better realize specific requirements. For example, it
is possible to choose a characteristic transition time (the time
needed to go from the initial to the final state) by inserting
exponential relaxation terms in the interpolating functions.

We will now supply a few explicit examples of transitions. Our
initial ($s\ll\tau$) Gaussian, coherent, oscillating wave function
has the form
\begin{equation}\label{initial}
  \psi_1(\xi,s)=\left(\frac{1}{2\pi\nu_1}\right)^{1/4}\,
  \exp\left[\frac{-(\xi-a_1\cos\omega_1s)^2}{4\nu_1}\,
            -i\,\frac{4a_1\xi\sin\omega_1s-a_1^2\sin2\omega_1s
               +4\nu_1\omega_1s}{8\nu_1}\right]\,,
\end{equation}
where we must also remember that
\begin{equation}\label{relation}
  \omega_1=\frac{\cal E}{\nu_1}\,.
\end{equation}
The relation~(\ref{relation}) means that our initial potential is
purely harmonic with frequency $\omega_1$. By
comparing~(\ref{initial}) with the required Gaussian density, i.e.
imposing that
\begin{equation}\label{density}
  |\psi_1(\xi, s)|^2=\rho(\xi,s)=
  \frac{{\mathrm e}^{-(\xi-\mu(s))^2/2\nu(s)}}{\sqrt{2\pi\nu(s)}}
\end{equation}
we get the initial identification
\begin{equation}\label{identification}
  \mu(s)=a_1\cos\omega_1s=a_1\cos\left(\frac{{\cal
  E}s}{\nu_1}\right)\,,\qquad\quad\nu(s)=\nu_1\,,\qquad\quad(s\ll\tau)\,.
\end{equation}
As for the initial phase function, by inspection of
equations~(\ref{initial}) and~(\ref{ansatz}), and by
taking~(\ref{relation}) into account, we immediately get
\begin{equation}\label{phasefunction}
  S(\xi,s)=m\omega_1\left(\frac{a_1^2}{4}\sin2\omega_1s-
  {\cal E}s-a_1\xi\sin\omega_1s\right)\,,\qquad\quad(s\ll\tau)\,.
\end{equation}

First of all we want to describe the (smooth) transition of our
initial wave function to a final one of the same form but
characterized by a new set of parameters:
\begin{equation}\label{transition}
  a_1\rightarrow
  a_2\,,\qquad\quad\nu_1\rightarrow\nu_2\,,\qquad\quad
  \omega_1=\frac{\cal E}{\nu_1}\rightarrow\omega_2=\frac{\cal
  E}{\nu_2}\,.
\end{equation}
The choice~(\ref{transition}) means that also the final potential
is still purely harmonic, but with a new frequency $\omega_2$. In
order to achieve that we consider for example the function
\begin{equation}\label{transitionfunction}
  \Gamma(s)=\frac{1}{1+{\mathrm e}^{-(s-\tau)/\gamma}}
\end{equation}
which smoothly goes from $0$ (for $s\ll\tau$) to $1$ (for
$s\gg\tau$) with a flex point in $s=\tau$ and a transition
velocity equal to $1/\gamma$. Of course here $\tau$ and $\gamma$
are completely free parameters: a suitable choice of them will
allow to fine tune the timing and the velocity of the transition.
Now the required transition is implemented by choosing
\begin{equation}\label{interpolation}
  \mu(s)=a_1\cos\left(\frac{{\cal E}s}{\nu_1}\right)(1-\Gamma(s))+
  a_2\cos\left(\frac{{\cal E}s}{\nu_2}\right)\Gamma(s)\,,\qquad
  \nu(s)=\nu_1(1-\Gamma(s))+\nu_2\Gamma(s)\,,
\end{equation}
which realizes~(\ref{transition}) and hence interpolates between
the two initial and final Gaussian, coherent, oscillating states.
\begin{figure}
\begin{center}
\includegraphics*[width=9cm]{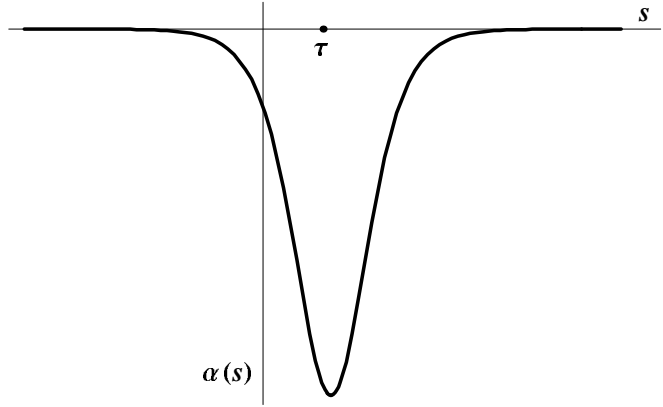}
\end{center}
\caption{The function $\alpha(s)$ is the coefficient of the
$\xi^2$ term in the phase function~(\ref{interpolatingphase}).
Notice that it goes quickly to zero, as required, outside the
transition region of width $\gamma$ around $s=\tau$. Its negative
values are due to the choice of monotonically decreasing
dispersion $\nu$ (better collimation).}\label{fig01}
\end{figure}
\noindent The phase function can now be calculated
from~(\ref{oscillating}) and we have
\begin{eqnarray}\label{interpolatingphase}
  S(\xi,s)&=&m\left[\alpha(s)\xi^2+\beta(s)\xi+H(s)+\theta(s)\right]\\
  \alpha(s)&=&\frac{\nu'}{4\nu}\,,\qquad\beta(s)=\mu'-\frac{\mu\nu'}{2\nu}\,,
  \qquad H(s)=\frac{\nu'\mu^2}{4\nu}\,.
\end{eqnarray}
Since $\alpha$, $\beta$ and $H$ are now fixed
by~(\ref{interpolation}), a comparison
between~(\ref{interpolatingphase}) and~(\ref{phasefunction}), and
in particular between the asymptotic ($s\rightarrow\pm\infty$)
expressions of the $\xi$--independent term of the phase, will
suggest the following form for the arbitrary $\theta(s)$ function:
\begin{equation}\label{theta}
  \theta(s)=\left[\frac{{\cal E}a_1^2}{4\nu_1}
            \sin\left(\frac{2{\cal E}s}{\nu_1}\right)
            -\frac{{\cal E}^2s}{\nu_1}\right](1-\Gamma(s))
           +\left[\frac{{\cal E}a_2^2}{4\nu_2}
            \sin\left(\frac{2{\cal E}s}{\nu_2}\right)
            -\frac{{\cal E}^2s}{\nu_2}\right]\Gamma(s)-H(s)\,.
\end{equation}
Finally the potential will have the form
\begin{eqnarray}\label{interpolatingpotential}
  V_c(\xi,s)&=&m\left[\frac{1}{2}G(s)\xi^2-F(s)\xi+W(s)\right]\\
  G(s)=\frac{{\cal E}^2}{\nu^2}-\frac{\nu''}{2\nu}
  +\frac{\nu'^2}{4\nu^2}\,,\qquad F(s)&=&\mu''+\mu G\,,\qquad
  W(s)=\frac{G\mu^2}{2}-\frac{\mu'^2}{2}-\frac{{\cal
  E}^2}{\nu}-\theta'(s)\,,
\end{eqnarray}
where now all the terms are given by the previous relations.
\begin{figure}
\begin{center}
\includegraphics*[width=9cm]{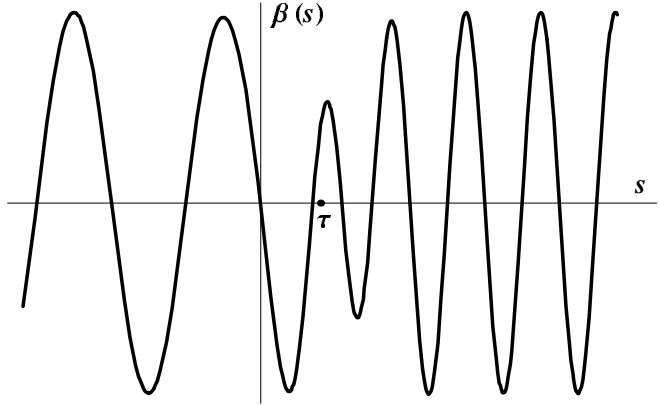}
\end{center}
\caption{The function $\beta(s)$ is the coefficient of the $\xi$
term in the phase function~(\ref{interpolatingphase}). Outside the
transition region it oscillates, as required, with stable
frequencies: $\omega_1$ for $s\ll\tau$, and $\omega_2$ for
$s\gg\tau$. The faster oscillation for $s\gg\tau$ is due to the
fact that $\omega_2>\omega_1$.}\label{fig02}
\end{figure}
\noindent As already remarked this potential has exactly the
form~(\ref{harmonic}). \noindent The functions $\alpha(s)$,
$\beta(s)$, $G(s)$, $F(s)$ and $W(s)$, which determine the
potential, can now be explicitly calculated for our example from
the equations~(\ref{interpolation}). Their analytic expressions
are by far too long (albeit elementary), however their graphical
behaviour is very simple and can be easily plotted. In particular
see the Figures 1 - 5 for a few typical diagrams displaying the
principal characteristics of these parameters which completely
define the transition.
\begin{figure}
\begin{center}
\includegraphics*[width=9cm]{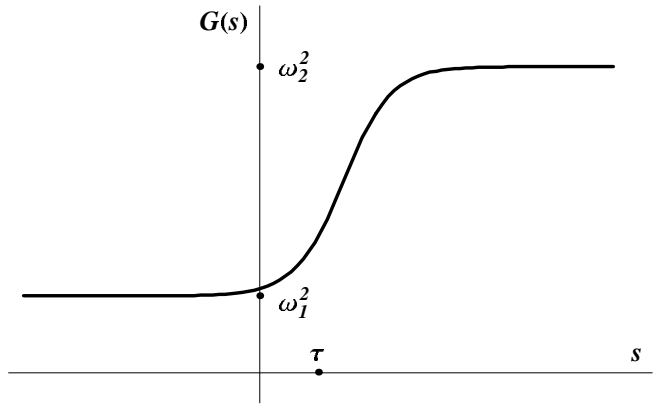}
\end{center}
\caption{The function $G(s)$ represents the square of the
time-dependent frequency of the harmonic, controlling
potential~(\ref{interpolatingpotential}).  Reminding
equation~(\ref{transition}) and that $\nu_2<\nu_1$ we have
$\omega_2>\omega_1$.}\label{fig03}
\end{figure}
\noindent First of all the functions $\alpha$ and $\beta$ show the
behaviour of the phase function: remark that it is not necessary
to produce a plot for the $\xi$--independent part of the phase
since the relation~(\ref{theta}) by definition imposes the right
asymptotic behaviour. Figure~\ref{fig01} shows that $\alpha(s)$
has a smooth extremal value around the transition at $\tau$, while
it also quickly goes to zero for $s\ll\tau$ and $s\gg\tau$: hence
no terms depending on $\xi^2$ remain asymptotically in the phase
as required by the form~(\ref{initial}). On the other hand
Figure~\ref{fig02} shows that $\beta(s)$ asymptotically has a
sinusoidal behaviour with different amplitudes and frequencies in
the two zones $s\ll\tau$ and $s\gg\tau$: this also is in good
agreement with the required form of the phase.
\begin{figure}
\begin{center}
\includegraphics*[width=9cm]{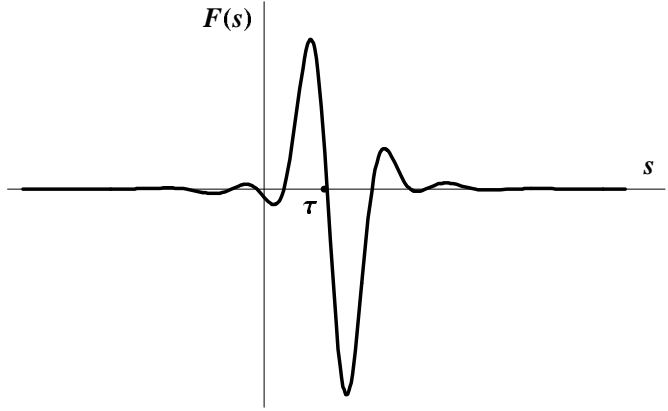}
\end{center}
\caption{The function $F(s)$ represents the time-dependent
coefficient of the $\xi$ term in the harmonic, control
potential~(\ref{interpolatingpotential}). The fact that it quickly
goes to zero outside the transition region is a consequence of the
relation $\omega_{2} = {\cal E}/\nu_{2}$ and of the
choice~(\ref{theta}) for the function $\theta(s)$.}\label{fig04}
\end{figure}
\noindent As for the control potential, Figure~\ref{fig03}
indicates that $G(s)$, which represents the parameter of the
harmonic part (depending on $\xi^2$) of $V_c$, smoothly goes from
$\omega_1^2$ to $\omega_2^2$ along the transition and sticks to
these two constant values outside the transition zone. From
Figure~\ref{fig04} and Figure~\ref{fig05} we finally see that
$F(s)$ and $W(s)$, which are respectively the coefficient of the
linear part and of the $\xi$-independent term in the control
potential, are different from zero only around the transition at
$s=\tau$, while they are everywhere zero far away from $\tau$. As
a consequence also the potential $V_c$ has the required time
behaviour since it is a simple harmonic potential for $s\ll\tau$
and $s\gg\tau$ (albeit with two different frequencies), and shows
some extra terms only in a limited interval around the transition.
\begin{figure}
\begin{center}
\includegraphics*[width=9cm]{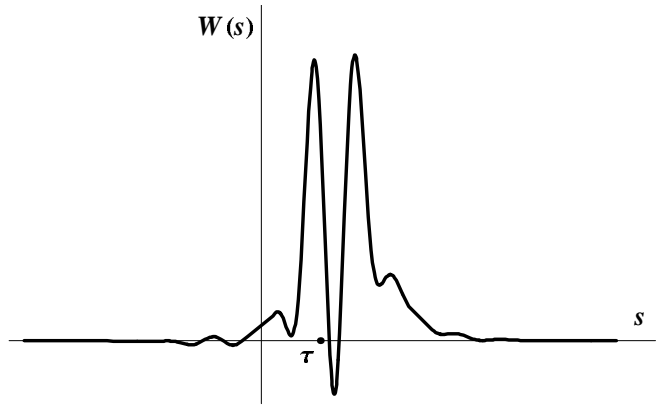}
\end{center}
\caption{The function $W(s)$ represents the coefficient of the
$\xi$-independent term in the harmonic, control
potential~(\ref{interpolatingpotential}). Here too the fact that
it quickly goes to zero outside the transition region is a
consequence of the relation $\omega_{2} = {\cal E}/\nu_{2}$ and of
the choice~(\ref{theta}) for the function
$\theta(s)$.}\label{fig05}
\end{figure}
\noindent Of course this does not constitute the only potential we
can obtain by this way. For example the function $\mu(s)$,
instead, could be chosen in such a way that the oscillation of the
centre of the profile be slower than the initial one, despite the
fact that the better collimation requires a final potential
associated to a frequency $\omega_{2} = {\cal E}/\nu_{2}$ larger
than the initial one and then to a faster betatron oscillation
with the same amplitude. This can be achieved by keeping a
suitable forcing part $F(s)$ different from zero also for
$s\gg\tau$: namely in this case the final potential does not
reduces itself to a simple harmonic one. It is easy to show that
if the final oscillation has the generalized form
\begin{equation}\label{newoscillation}
\mu (s) = a \,\cos (\omega s) + \frac{b}{m} \, \sin (\omega s),
\end{equation}
with $\omega$ not coincident with ${\cal E}/\nu$, the final
forcing function $F(s)$ calculated
from~(\ref{interpolatingpotential}) will correspondingly be
\begin{equation}\label{forcing}
F(s) = m \left(\omega^2  - \frac{{\cal E}^2}{\nu^2}\right) \left(a
\, \cos \omega s + \frac{b}{m} \, \sin \omega s\right).
\end{equation}
\begin{figure}
\begin{center}
\includegraphics*[width=9cm]{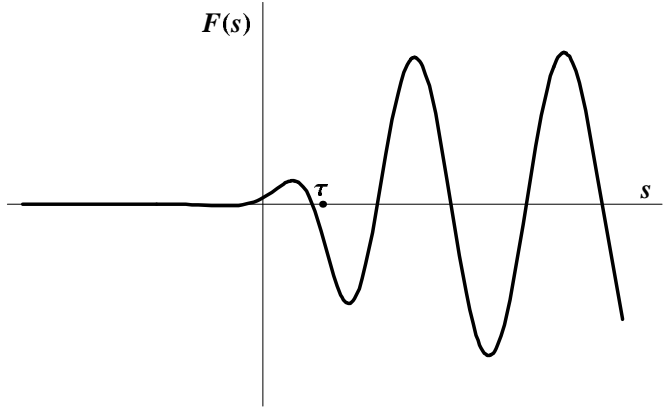}
\end{center}
\caption{This behaviour of the function $F(s)$, different from
that of Figure~\ref{fig04}, is due to the fact that the relation
$\omega_{2} = {\cal E}/\nu_{2}$ is no longer satisfied and the
choice~(\ref{newtheta}) is taken for the function $\theta(s)$. The
non vanishing oscillations of $F(s)$ in the asymptotic region
$s\gg\tau$ allow to reduce the otherwise naturally enhanced
betatron oscillations.}\label{fig06}
\end{figure}
\noindent In this case the potentials are more complicated but can
still be suitably explored by means of our method. As an example
we consider the case where the final state is characterized by two
independent parameters: $\omega_2$ for the frequency and $\nu_2$
for the packet spreading. Now a relation similar
to~(\ref{relation}) will be no longer satisfied. As a consequence
the choice~(\ref{interpolation}) will be changed in
\begin{equation}\label{newinterpolation}
  \mu(s)=a_1\cos\left(\frac{{\cal E}s}{\nu_1}\right)(1-\Gamma(s))+
  a_2\cos\left(\omega_2 s\right)\Gamma(s)\,,\qquad
  \nu(s)=\nu_1(1-\Gamma(s))+\nu_2\Gamma(s)\,,
\end{equation}
while we get a new determination for the arbitrary $\theta(s)$
function:
\begin{equation}\label{newtheta}
  \theta(s)=\left[\frac{{\cal E}a_1^2}{4\nu_1}
            \sin\left(\frac{2{\cal E}s}{\nu_1}\right)
            -\frac{{\cal E}^2s}{\nu_1}\right](1-\Gamma(s))
           +\left[\frac{\omega_2 a_2^2}{4}
            \sin\left(2\omega_2 s\right)
            -{\cal E}\omega_2 s\right]\Gamma(s)-H(s)\,.
\end{equation}
\begin{figure}
\begin{center}
\includegraphics*[width=9cm]{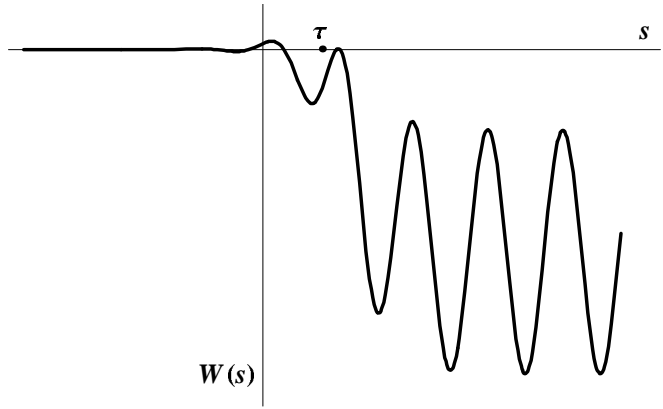}
\end{center}
\caption{Here too, as for the function $F(s)$, the new behaviour
of the function $W(s)$, different from that of Figure~\ref{fig05},
is due to the fact that the relation $\omega_{2} = {\cal
E}/\nu_{2}$ is no more satisfied and the choice~(\ref{newtheta})
is taken for the function $\theta(s)$.}\label{fig07}
\end{figure}
\noindent The functions defining the time evolution of both the
phase and the potential can now be calculated once more and we
find that, the functions $\alpha(s)$, $\beta(s)$ keep a form very
similar to the previous one. Instead the new $G(s)$ displays an
opposite behaviour with respect to the Figure~\ref{fig03}: in this
case the final frequency $\omega_2$ is smaller than the initial
frequency $\omega_1$, and thus the betatron oscillations are
suppressed. On the other hand the shape is still of the form of a
sigmoid. As for the functions $F(s)$ and $W(s)$ they show a
different asymptotic behaviour as can be seen from
Figure~\ref{fig06} and~\ref{fig07}.
\begin{figure}
\begin{center}
\includegraphics*[width=9cm]{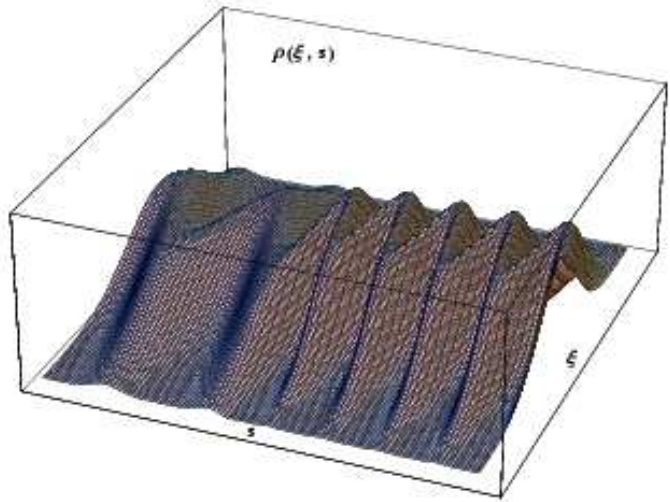}
\end{center}
\caption{The density~(\ref{density}) of the bunch as a function of
$s$ and $\xi$. Here the squeezing is performed with the
constrains~(\ref{transition}) between the parameters $\omega$ and
$\nu$, and hence without control on the betatron oscillations.
Hence these oscillations are enhanced as a consequence of the
squeezing.}\label{fig08}
\end{figure}
\noindent In particular we see that, as predicted, $F(s)$ and
$W(s)$ no more disappear for $s\gg\tau$, so that asymptotically we
do not have a purely harmonic potential since now
in~(\ref{harmonic}) both the term linear and that constant in
$\xi$ will be present for every $s>\tau$. However it is clear that
other choices are always possible: for example the arbitrary
function $\theta(s)$ could be defined so that
in~(\ref{interpolatingpotential}) the $\xi$-independent term
$W(s)$ of the potential $V_c$ be identically zero. Of course there
would be a price to pay for that: in fact now in the phase
function $S$ the $\xi$-independent term will no more follow an
asymptotic behaviour of the type~(\ref{phasefunction}) since the
relation~(\ref{newtheta}) will no more be satisfied.
\begin{figure}
\begin{center}
\includegraphics*[width=9cm]{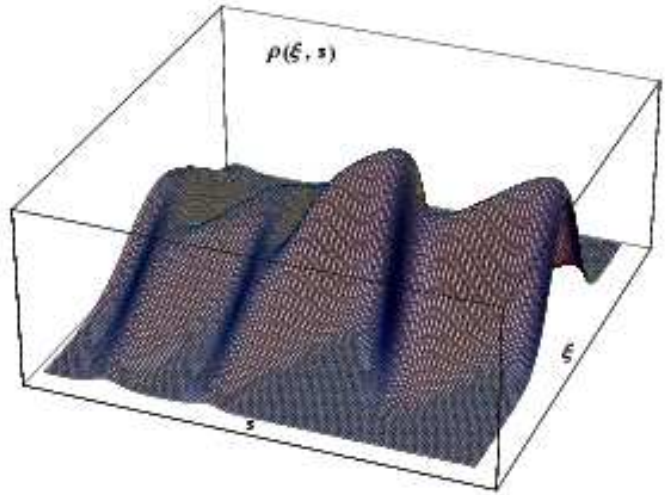}
\end{center}
\caption{At variance with Figure~\ref{fig08}, here the
density~(\ref{density}) is squeezed with no constrains of the
type~(\ref{transition}) between the parameters $\omega$ and $\nu$.
As a consequence we were able to slow down at the same time also
the betatron oscillations.}\label{fig09}
\end{figure}
\noindent In the most general case of transitions between states
with non constant dispersion (strong focusing) it is clear that
the procedure can also be suitably extended. In fact it is
sufficient to exploit for instance the
expression~(\ref{interpolation}) for the interpolating dispersion,
but with time dependent initial and final dispersions $\nu_{1}
(s)$ and $\nu_{2} (s)$. The general
form~(\ref{interpolatingpotential}) of the controlling potential
is thus calculated, but with a new expression for $\nu (t)$.
Finally, also the initial and final laws of motion of the profile
centre, $\mu_{1} (s)$ and $\mu_{2} (s)$, can always be chosen as
in the previously discussed example. However, in this case, a
forcing part $F(s)$  is needed to retain the oscillatory
motion~(\ref{newoscillation}) for $s\gg\tau$. In conclusions
Figures~\ref{fig08} and~\ref{fig09} show the $s$--evolution of the
density of the bunch. Both describe a squeezing of the beam, but
Figure~\ref{fig08} reproduces the case where the frequency of the
betatron oscillation is enhanced, while Figure~\ref{fig09} is
related to the case where these oscillations are reduced.

\section{Conclusions}\label{conclusions}

In the first part of this paper we have applied to the collective
dynamics of beams in particle accelerators a stability analysis
already developed for general particle systems. This analysis has
allowed us to single out scaling factors relating the parameters
ruling the collective dynamics in the beams with the microscopic
scales.

In the second part of the paper we have considered the stability
regime of a beam, in which the energy loss due to the radiation
damping is on average compensated by the external RF energy
pumping. The collective beam dynamics in this regime is described
by time--reversal invariant diffusion processes (Nelson processes)
which are obtained by a stochastic extension of the least action
principle of classical mechanics. The choice of the diffusion
coefficient is dictated by the unit of emittance determined in the
first part of the paper. The collective dynamics of beams is then
described by two non-linearly coupled hydrodynamic equations. It
has also been observed that the linearization of these equations
connects this approach to a Schr\"odinger--like (quantum--like)
effective description of the beam dynamics previously developed
through different approaches.

In the last part of the paper we have shown that the transition
probabilities of Nelson processes can be exploited to control the
collimation and the oscillations of the beam in the quadrupole
approximation, both in the weak focusing and in the strong
focusing regimes. In this framework we have explicitly computed
the controlling potentials that realize some relevant controlled
evolutions. The controlling potentials can be engineered by
suitable tuning of the external RF and magnetic fields. We have
considered evolutions that drive the beam from a less collimated
to a better collimated state. We have furthermore shown that this
goal can also be achieved without increasing the frequency of the
betatron oscillations which can in fact be independently
controlled during the evolution. In the forthcoming papers we will
study the extension of these control techniques beyond the
quadrupole approximation and address in detail applications to
existing machines, problems related to dynamical instabilities and
topics about the halo formation, a problem which has recently been
addressed in the framework of a quantum--like
approach~\cite{pusterla}.

\appendix
\section{Minimal action and scaling factors for
stable systems}\label{minimalaction}

We consider a generic stable system confined in a region of space
of linear dimension $R$, constituted by a large number $N$ of
identical particles of mass $m$, and ruled by an attractive
classical (possibly effective) law of force $F(r)$. We introduce a
unit of action $\alpha$ (which will turn out to be minimal) by the
following relation:
\begin{equation}\label{appendixaction}
\alpha=m {\tilde v}^{2}\tau.
\end{equation}
In this equation, ${\tilde v}$ denotes the characteristic mean
velocity per particle in the system, while $\tau$ is a
characteristic microscopic time whose size must be
self--consistently determined (for details see
reference~\cite{demartino}). In order to obtain an explicit
expression for $\alpha$, we then impose the following criteria of
stability: firstly, we require that the characteristic potential
energy of each particle be on average equal to its characteristic
kinetic energy (virial theorem):
\begin{equation}\label{virial}
{\cal{L}} \cong m{\tilde v}^{2},
\end{equation}
where $\cal{L}$ is the work performed in mean by the entire system
on a single constituent. Then, if the system extends on the
characteristic length scale $R$ we have, in order of magnitude
\begin{equation}\label{work}
{\cal{L}} \cong NF(R)R,
\end{equation}
where $F(R)$ is the force evaluated on a distance of the order of
magnitude of the linear global dimension of the system.
Relations~(\ref{virial}) and~(\ref{work}) are now summarized by
the following expression of the characteristic velocity $\tilde
v$:
\begin{equation}\label{appendixvelocity}
{\tilde v} \cong \sqrt{ \frac{N F(R)R}{m}}.
\end{equation}
We then define the macroscopic time scale ${\cal{T}}$ associated
to the entire system, through the obvious relation ${\tilde
v}=R/{\cal{T}}$ (therefore ${\cal{T}}$ has the meaning of a
characteristic traveling time for a particle inside the system).
We insert into equation~(\ref{appendixaction}) both the latter
expression and the expression~(\ref{appendixvelocity}), obtaining
the following form for the (minimal) unit of action:
\begin{equation}\label{minimalaction2}
\alpha \cong \sqrt{m F(R)} R^{3/2} N^{1/2} \frac{\tau }{\cal{T}}.
\end{equation}
We now introduce a second requirement for the mechanical
stability, namely that due to the large number of particles, the
unit of action $\alpha$ be not sensibly dependent on $N$. As a
natural consequence, we are led to impose a relation between the
microscopic characteristic time $\tau$ and the macroscopic
characteristic traveling time ${\cal{T}}$ of the form
\begin{equation}\label{time}
\tau \cong \frac{{\cal{T}}}{\sqrt{N}}.
\end{equation}
Finally, by inserting~(\ref{time}) into
equation~(\ref{minimalaction2}) we finally obtain the (minimal)
unit of action
\begin{equation}\label{appendixunit}
\alpha\cong m^{1/2}R^{3/2} \sqrt{F(R)}.
\end{equation}

\end{document}